\documentclass[11pt,preprint]{emulateapj}
\usepackage{amsmath,euscript,amsfonts}
\def\cfa{1}
\def\msu{2}
\def\nrao{3}
\def\jansky{4}
\def\ukansas{5}
\def\itp{6}
\def\uv{7}
\def\ioa{8}
\def\su{9}
\def\Tau{10}

\shorttitle{The Radio Luminous Hypernova 2012au}
\shortauthors{Kamble et al.}

\begin{document}

\title{A Wolf in Sheepskin:\\ Extraordinary Supernova 2012au Veiled Behind Ordinary Radio Emission}

\author{
Atish Kamble\altaffilmark{\cfa}, 
Alicia ~M. Soderberg\altaffilmark{\cfa},
Laura Chomiuk\altaffilmark{\msu,\nrao,\jansky},
Raffaella Margutti\altaffilmark{\cfa}, 
Mikhail Medvedev\altaffilmark{\ukansas,\itp},
Sayan Chakraborti\altaffilmark{\cfa},
Roger Chevalier\altaffilmark{\uv},
Nikolai Chugai\altaffilmark{\ioa},
Jason Dittmann\altaffilmark{\cfa},
Maria Drout\altaffilmark{\cfa},
Claes Fransson\altaffilmark{\su},
Dan Milisavljevic\altaffilmark{\cfa},
Ehud Nakar\altaffilmark{\Tau}, 
Nathan Sanders\altaffilmark{\cfa}
}

\altaffiltext{\cfa}{Harvard-Smithsonian Center for Astrophysics,
	60 Garden St., Cambridge, MA 02138}

\altaffiltext{\msu}{Department of Physics and Astronomy, Michigan State University, 
	East Lansing, MI 48824, USA}

\altaffiltext{\nrao}{National Radio Astronomy Observatory, P.O. Box O, Socorro, NM 87801, USA}

\altaffiltext{\jansky}{Jansky Fellow}

\altaffiltext{\ukansas}{the Department of Physics and Astronomy, University of Kansas,
	Lawrence, KS 66045, USA}

\altaffiltext{\itp}{the ITP, NRC ÒKurchatov Institute,Ó Moscow 123182, Russia}

\altaffiltext{\uv}{Department of Astronomy, University of Virginia, P.O. Box 400325, 
	Charlottesville, VA 22904-4325, USA}

\altaffiltext{\ioa}{Institute of Astronomy, Russian Academy of Sciences, 
	Pyatnitskaya 48, 109017 Moscow, Russia}

\altaffiltext{\su}{Department of Astronomy, The Oskar Klein Centre, 
	Stockholm University, AlbaNova University Centre, SE-106 91 Stockholm, Sweden}

\altaffiltext{\Tau}{Raymond and Beverly Sackler School of Physics and 
	Astronomy, Tel Aviv University, Tel Aviv 69978, Israel}

\begin{abstract}
	We present extensive radio and X-ray observations of SN\, 2012au, the energetic 
	radio luminous supernova of type Ib that may be a link between subsets of hydrogen-poor 
	superluminous and normal core-collapse supernovae.
	The observations closely follow models of synchrotron emission from shock heated 
	circum-burst medium that has a wind density profile ($\rho \propto r^{-2}$). 
	We infer a sub-relativistic 
	velocity for the shock wave $v \approx 0.2\,c$ and a radius of $r \approx 1.4 \times 
	10^{16}~\rm cm$ at 25 days after the estimated date of explosion. For a constant 
	wind velocity of $1000$ km/s we determine the constant mass loss rate 
	of the progenitor to be $\dot{M} = 3.6 \times 10^{-6} ~\rm M_{\odot} ~yr^{-1}$,
	consistent with the estimates from X-ray observations.
	We estimate the total internal energy of the radio emitting material to be 
	$E \approx 10^{47} ~\rm erg$, which is intermediate to SN\,1998bw and SN\,2002ap.
	Evolution of the radio light curves of SN\,2012au is consistent with interaction with a smoothly 
	distributed circum-burst medium and absence of stellar 
	shells ejected from previous outbursts out to $r \approx 10^{17} ~\rm cm$ 
	from the supernova site. 
	Based on this we conclude that the evolution of the SN\,2012au progenitor star 
	was relatively quiet during the final years preceding explosion. 
	We find that the bright radio emission from SN2012au was not dissimilar from 
	other core collapse supernovae despite it's extraordinary optical properties. 
	We speculate that it was the nature of the explosion that led to 
	the unusual demise of the SN2012au progenitor star. 
\end{abstract}

\keywords{
	radiation mechanisms: nonthermal - 
	radio continuum: general - supernovae: general 
	- supernovae: individual (SN\, 2012au)
}

\section{Introduction}
	Optical studies have recently discovered a new class of supernovae --
	Super Luminous Supernovae (SLSNe) with absolute magnitude of $\leq
	-21$ \citep{Quimby2011, GalYam2012}. Models of SLSNe suggest very
	massive progenitors $M>70 ~M_{\odot}$ which undergo dynamical
	instability due to the formation of $e^{-}-e^{+}$ pairs \citep{Barkat1967}. The recent
	discovery and follow-up data for SN\,2012au reveals several features
	of SLSNe in the optical light curve and spectral evolution, as shown
	by \citet{Milisavljevic2013,Takaki2013}, which suggests it to be a link
	in the chain : the normal core collapse supernovae $\rightarrow$
	hypernovae $\rightarrow$ SLSNe.  Such explosive events have been
	suggested to have pre-explosion pair-instability eruptions \citep{Chatzopoulos2012}
	creating shells of CSM in the vicinity of the SN progenitor system.
	Radio emission from supernovae serves as a probe of the circum-burst
	environment through the excitation of local particles to
	relativistic velocities \citep{Weiler2002} and may therefore be used
	to search for CSM shells. Although a few candidate SLSNe have been tried
	\citep{Chomiuk2011}, to date there has been no radio detection 
	of a SLSN. This makes SN\, 2012au a potential first look at the 
	mass loss properties of a SLSN.

	In this article, we report radio and X-ray observations and analysis of
	SN\,2012au, and show that it had a radio luminosity and associated energy intermediate
	between the broad-lined and energetic type Ic SNe 2002ap and
	1998bw. SN\,2012au points to the existence of a population of supernovae
	occupying the phase space between normal and relativistic core
	collapse supernovae. The organization of this paper is as follows:
	Radio observations using the Jansky Very Large Array (VLA) and X-ray observations from
	\emph{Swift} X-ray Telescope (XRT) are described
	in Section \ref{sec:obs-rad} and \ref{sec:obs-xray},
	respectively. Preliminary estimates of the physical parameters
	governing the explosion and radiation, based on the assumption of
	equipartition, are obtained in Section \ref{sec:equipartition}.
	In Section \ref{sec:ssa} we present detailed analysis and
	resultant parameters without the equipartition assumption. 
	The inverse Compton X-ray emission from SN\,2012au is discussed 
	in Section \ref{sec:IC} and then we try to constrain the microphysical 
	parameters $\epsilon_{e}$ and $\epsilon_{B}$ in Section \ref{sec:constraints}.
	We discuss
	implications of SN\,2012au in regards to different models such as Pair
	Instability Supernovae (PISN) and Pulsational Pair Instability
	Supernovae (PPISN) (\ref{sec:pisn}),
	and compare it with other energetic SNe Ib/c (\ref{sec:hypern})
	and also the possibility of it being asymmetric (\ref{sec:asymme}).
	Comparison with other SNe in energy-velocity phase space is discussed in Section 
	\ref{sec:EkGammaBeta}. Our conclusions are summarized in Section \ref{sec:conclu}.

\section{Observations}
\label{sec:obs}
	
	\subsection{Radio Observations using \emph{Very Large Array}}
	\label{sec:obs-rad}
	
	SN\,2012au was optically discovered on 14 March 2012 by the CRTS SNHunt project,
	with an offset of about $3^{\prime\prime}.5$ E and $2^{\prime\prime}.0$ N from 
	the center of the host galaxy NGC 4790 \citep{Howerton2012} at a distance of
	$d\approx 23.5$ Mpc \citep{Theureau2007}.  In our first observation with the
	VLA \footnote{The Jansky Very Large Array 
	is operated by the National Radio Astronomy Observatory, a
	facility of the National Science Foundation operated under cooperative
	agreement by Associated Universities, Inc.} on 2012 March
	20.2 UT, we detected a bright radio source coincident with the optical
	position at $\alpha\rm (J2000)=12^{\rm h} 54^{\rm m} 52.18^{\rm s}$ and
	$\delta\rm (J2000)=-10^{\rm o}14^{\prime}50.2^{\prime\prime}$ ($\pm$0.1 arcsec in each
	coordinate) with flux density of $f_{\nu}=4.59\pm 0.025~m$Jy at 5 GHz.
	This, and subsequent observations from 5 to 37 GHz, are
	summarized in Table~1.
	
	All observations were taken in standard continuum observing mode with
	a bandwidth of $16\times 64$ MHz. During the reduction
	we split the data in two side bands ($8\times 64$ MHz) of
	approximately 1 GHz each. We used 3C286 for flux calibration,
	and for phase referencing we used calibrators J1305-1033. 
	Data were reduced using standard packages within the
	Astronomical Image Processing System (AIPS).
	
\begin{figure}
		\begin{center}
			\includegraphics[width=9cm,clip=False]{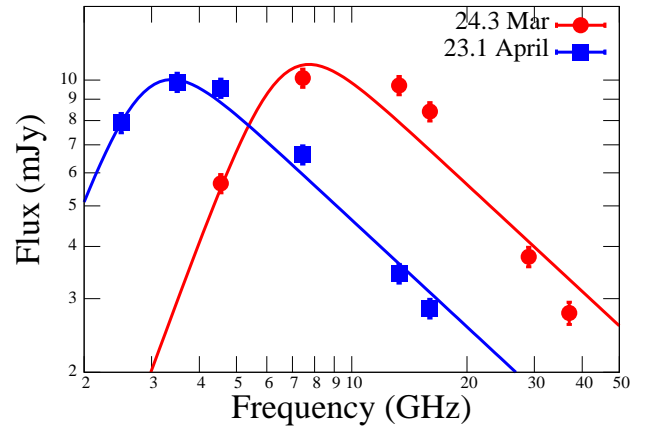}
			\caption{Radio SED of SN\,2012au on 24.3 March and 23.1 April 2012, 
			about 21 and 51 days after the supernova, respectively.}
			\label{fig:sed}
		\end{center}
	\end{figure}

\begin{figure*}
		\begin{center}
			\includegraphics[scale=0.7]{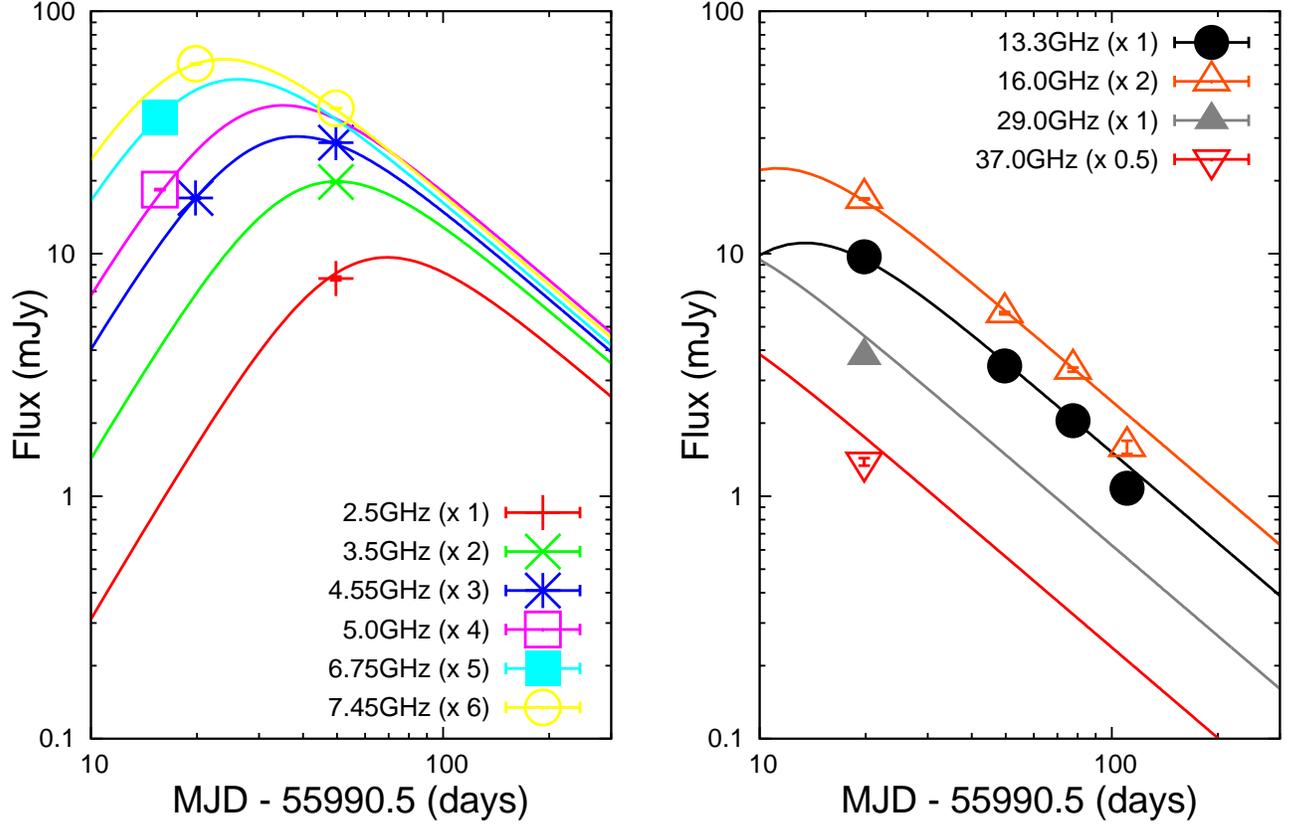}
			\caption{Radio light curves of type Ic SN\, 2012au as observed with the VLA at frequencies 
			from 2.5 GHz to 37 GHz between March 20 and June 23, 2012. The model light curves, 
			i.e. synchrotron self-absorbed radiation from the non-relativistic freely expanding blast wave 
			interacting with the wind stratified medium, have been over plotted the data points.
			Both, the model light curves and the data points, have been off-set for the purpose of clarity.}
			\label{fig:lc}
		\end{center}
	\end{figure*}

	The radio spectral energy distribution (SED) of SN\,2012au as 
	observed on March 24 and April 23, 2012 are shown in Figure~\ref{fig:sed}. 
	In the optically thin regime we measure $f_{\nu}\propto \nu^{-0.85\pm0.02}$.
	Figure~\ref{fig:lc} shows the SN\,2012au radio light-curves
	assuming an approximate explosion date of 2012 March 3.5 UT
	(MJD  55990.5 $\pm$ 2.0)  following \citet{Milisavljevic2013}. 
	
	\begin{deluxetable}{lcccr}
	\tablecaption{VLA radio flux density measurements of SN\,2012au}
	\tablewidth{0pt}
	\tablehead{ 
		\colhead{Date}	& \colhead{MJD}	&	\colhead{Frequncy}	& \colhead{$F\pm \sigma$}	& \colhead{Array} \\
		\colhead{(UT)} 				& \colhead{} 		& 	\colhead{(GHz)} 	& \colhead{(mJy)} 		& \colhead{Config.}
	}
	\startdata
		2012 Mar 20.2	&  	56006.2	&	5.0	&	4.596  $\pm$ 0.025	&	C		\\
		\ldots               	&	\ldots	&	6.75	&      7.294  $\pm$ 0.019	&	\ldots	\\
		2012 Mar 24.3	&	56010.3	& 	4.55 	&  	5.658  $\pm$  0.036 &	C		\\
		\ldots               	&	\ldots	&	7.45	&    	10.113$\pm$  0.020 &	\ldots	\\
		\ldots               	&	\ldots	&	13.3	&    	9.709  $\pm$ 0.049 	&	\ldots	\\
		\ldots		&      \ldots	&	16.0	&    	8.408  $\pm$ 0.058	&	\ldots	\\ 
		\ldots		&      \ldots	&	29.0	&    	3.778  $\pm$ 0.072 	&	\ldots	\\
		\ldots        		&      \ldots	&	37.0	&    	2.772  $\pm$ 0.098 	&	\ldots	\\
		2012 Apr 23.1	& 	56040.1	&	2.5	&     	7.900  $\pm$ 0.142 	&	C		\\
		\ldots               	&	\ldots	&	3.5	&     	9.886  $\pm$ 0.053 	&	\ldots	\\
		\ldots               	&	\ldots	&	4.55	&    	9.569  $\pm$ 0.052 	&	\ldots	\\
		\ldots               	&	\ldots	&	7.45	&    	6.630  $\pm$ 0.024 	&	\ldots	\\
		\ldots               	&	\ldots	&	13.3	&    	3.446  $\pm$ 0.040 	&	\ldots	\\
		\ldots               	&	\ldots	&	16.0	&    	2.843  $\pm$ 0.034 	&	\ldots	\\
		2012 May 21.1	&	56068.1	&	13.3	&    	2.049  $\pm$ 0.034	&	CnB		\\
					&	\ldots	&	16.0	&    	1.662  $\pm$ 0.032	&	\ldots	\\
		2012 Jun 23.0	& 	56101.0	&	13.3	&	1.079 $\pm$ 0.049	&	B		\\
		\ldots		& 	\ldots	&	16.0	&	0.797 $\pm$ 0.050	&	\ldots	\\
	\enddata
	\label{tab:vla}
	\end{deluxetable}
	
	\subsection{X-ray Observations using \emph{Swift} \rm}
	\label{sec:obs-xray}
	
	We observed SN\,2012au with the \emph{Swift} \citep{Gehrels04} X-Ray Telescope 
	(XRT, \citealt{Burrows05}) starting from 2012 March 15 until 2012 April 21, or roughly
	between 12 and 48 days since the supernova. 
	Data have been reduced with HEAsoft v. 6.13 and corresponding calibration files. 
	Standard filtering and screening criteria have been applied. Using a total of 34.7 ks, 
	we find evidence for significant X-ray emission at a position consistent with 
	SN\,2012au, with 0.3-10 keV count-rate of $(11.6\pm2.5)\times 10^{-4}\rm{c\,s^{-1}}$
	corresponding to a $4.6\,\sigma$ detection. Adopting a simple power-law spectrum 
	and correcting for the Galactic absorption in the direction of SN\,2012au 
	($\rm {NH_{MW}}=3.7\times 10^{20}\,\rm{cm^{-2}}$, \citealt{Kalberla05}), 
	the measured count-rate translates into an unabsorbed flux of 
	$(4.6\pm 1.0)\times 10^{-14}\,\rm{erg\,s^{-1}cm^{-2}}$ ($L_{X} 
	\sim 3\times 10^{39}\,\rm{erg\,s^{-1}}$).
	
	SN\,2012au is positioned at less than 10$^{\prime\prime}$ far from the center of NGC4790's 
	central nucleus (\citealt{Milisavljevic2013}, their Fig. 1). The Half Energy 
	Width (HEW) of \emph{Swift}-XRT is $\sim18^{\prime\prime}$
	(for an on-axis source at 1.5 keV, \citealt{Moretti05}) thus, X-ray emission from 
	the host nucleus might contaminate the detected X-ray emission. 
	However, splitting the data set into two halves with exposure times of 9.8 ks 
	and 24.9 ks, respectively, and applying a binomial test, we find evidence 
	for fading of the X-ray emission, with a probability of a chance fluctuation 
	of $0.7\%$. The detection of fading on a time-scale of $\sim 30$ days suggests 
	that SN\,2012au dominates the X-ray emission we detect. In the following 
	we proceed with this hypothesis. 
	
	\begin{figure}[h]
		\begin{center}
			\includegraphics[width=8cm]{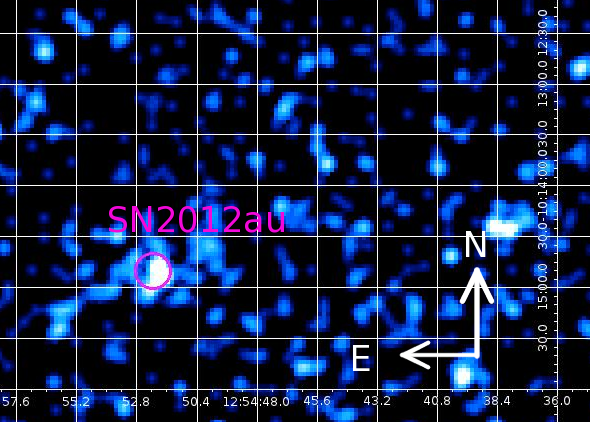}
			\caption{\emph{Swift} X-ray image of SN\,2012au}
			\label{fig:ximage}
		\end{center}
	\end{figure}

\section{Preliminary Constraints from Equipartition Assumption}
\label{sec:equipartition}
	Radio observations of supernovae primarily trace the shocked ejecta and have proven 
	to be a good probe 
	of the velocity of the shock wave and kinetic energy of the ejecta. To first degree approximation
	these quantities can be estimated quickly, without detailed observations, using the reasonable
	assumption of energy equipartition between magnetic field ($\epsilon_{B}$) and accelerated 
	electrons ($\epsilon_{e}$). Under the assumption 
	of equipartition ($\epsilon_{e} = \epsilon_{B} = 0.5$) the angular radius of the shock wave, 
	$\theta_{ep}$, and the equipartition energy, 
	$E_{ep}$, are given by

	\begin{align}
	\theta_{ep}&	\approx  
		120 ~\left( \frac{d_{L}}{\rm Mpc} \right)^{-1/17} \left( \frac{f_{\nu_{a}}}{\rm mJy} \right)^{8/17} 
		\left(\frac{ \nu_{a}}{\rm GHz} \right)^{-(36-p)/34}		~\mu{\rm as}	\label{eqn:equipartition1}	\\
	E_{ep} 	&	\approx 
		4 \times 10^{43} \left( \frac{d_{L}}{\rm Mpc} \right)^{40/17} \left( \frac{f_{\nu_{a}}}{\rm mJy} \right)^{20/17} 
		\left(\frac{ \nu_{a}}{\rm GHz} \right)^{-(3p+11)/17}	{\rm erg}	\label{eqn:equipartition2}	
	\end{align}
	following \citet{Chevalier1998,Kulkarni1998} and \citet{Soderberg2005}.
	Using our VLA observation of SN\,2012au on 24.3 March 2012 ($t=20.8$ days), 
	we find that the self-absorption frequency $\nu_{a} \approx 8$ GHz, corresponding flux 
	$f_{\nu_{a}} \approx 10$ mJy and $p\approx 2.7$. Using equations \ref{eqn:equipartition1} 
	and \ref{eqn:equipartition2}, 
	these parameters give energy $E_{ep} \approx 10^{47}$ erg and the angular size of the shock wave 
	$\theta_{ep} \approx 39 ~\mu$arc-sec. At a distance $d_{L} \approx 23.5$ Mpc, this size translates 
	to the equipartition 
	radius of $r_{ep} \approx 1.4 \times 10^{16}$ cm and the expansion velocity $v\equiv\beta\,c \approx 0.26\,c$.
	Most core-collapse supernovae have $\beta \approx 0.1-0.15$ (see e.g. 
	\citealt{Chevalier2006,Soderberg2010}) which makes this a relatively faster supernova shock wave.
	
\section{Radio Luminosity and Synchrotron Self-Absorption}
\label{sec:ssa}
	
	Although equipartition arguments provide reasonable estimates of the important physical parameters 
	that drive the explosion and its evolution, it is far from clear if the equipartition holds in general 
	for supernova blast-waves or if it sets in at a particular stage during the evolution. Therefore,
	to do away with the assumption of equipartition and to treat the supernovae in general,
	in the Appendix we have developed a formulation of SN blast-wave driven synchrotron radiation 
	and its evolution following \citet{Chevalier1998,Soderberg2005,Chevalier2006}.
	
	Consider a shock wave expanding into the cirum-burst medium with velocity $\beta$, 
	expressed in units of the speed of light. The size and speed of the shock wave evolve 
	in time as
	\begin{eqnarray}
		r 	&	= 	&	r_{0} (t/t_{0})^{m}\\
		\beta &	= 	&	\beta_{0} (t/t_{0})^{m-1}
	\end{eqnarray}
	where $\beta c = dr/dt$ and the subscript `$0$' refers to quantities measured at epoch $t_{0}$.
	The circum-burst medium is assumed to be of wind profile, $n(r) \propto r^{-2}$.
	The synchrotron emission from SN blast-waves
	peak at radio frequency bands. The observed multiband spectral energy distribution due to a SN blast-wave 
	could be characterized by three observables: two break frequencies ($\nu_{m} $ and $\nu_{a}$) 
	and the spectral peak flux ($f_{\nu_{a}}$). As the break frequencies evolve in time, flux measured 
	at a given fixed frequency changes smoothly resulting in a light curve. The complete evolution of 
	the blast-wave could be described by four 
	independent physical parameters: expansion velocity of the blast wave $\beta$,
	fractional energy deposited in the relativistic electrons $(\epsilon_{e}) ~\&$ in the magnetic field 
	$(\epsilon_{B})$, and the
	circum-burst density into which the blast-wave is expanding $n(r)$ which could also be 
	written either as a wind parameter $A_{\star}$ or the mass loss rate of the star $\dot{M}$ 
	as defined in the Appendix. Also we define $\epsilon_{f} = \epsilon_{e}/\epsilon_{B}$. 
	The situation $\epsilon_{f}=1$ corresponds to equipartition 
	i.e., the equal amount of fractional energy in accelerated electrons and magnetic fields.
	Equations \ref{eqn:SpecPara1}, \ref{eqn:SpecPara2} and \ref{eqn:SpecPara3} 
	could then be inverted to get
	physical parameters of the explosion in terms of the observables.
	\begin{eqnarray}
		\beta_{0} 	&	= 	&
			0.18 ~m ~d_{L,100}^{16/17}
		   	~f_{\nu_{a},mJy}^{8/17} ~\nu_{a,10}^{-33/34}  ~\nu_{m,7}^{-1/34} t^{-m}_{0,d}
		   	~\epsilon^{-1/17}_{f}	\hfill \label{eqn:beta0}\\
		A_{\star} 	&	= 	&
			0.65 ~d_{L,100}^{22/17} ~f_{\nu_{a},mJy}^{11/17}
		   	~\nu_{a,10}^{29/34} ~\epsilon_{f}^{5/17} ~\nu_{m,7}^{-29/34} \hfill \label{eqn:astar0}\\
		\epsilon_{B}	&=	&
			0.29 ~m^{-2}~ \nu_{a,10}^{47/34}
		   	~\nu_{m,7}^{21/34} ~t_{0,d}^2 d_{L,100}^{-30/17}
		   	~f_{\nu_{a},mJy}^{-15/17} ~\epsilon_{f}^{-13/17} \hfill \label{eqn:epsilonB0}
	 \end{eqnarray}
	
	The numerical coefficients are the results of particular choice of units of the observables:
	$f_{\nu_{a}}$ in mJy, $\nu_{a}$ in $10^{10}$ Hz, $\nu_{m}$ in $10^{7}$ Hz, $t_{0}$ in days
	and distance $d_{L}$ in 100 Mpc.

	\subsection{Model Fits, Spectral and Physical Parameters}
	\label{sec:modeling}
		To fit all the radio observations collectively we used the smooth broken power law 
		\begin{equation}
			f_{\nu}(\nu,t) = f_{a,0} \times \left[ \left( \frac{\nu}{\nu_{a}(t)} \right)^{-\alpha_{1} s} 
			+ \left( \frac{\nu}{\nu_{a}(t)} \right)^{-\alpha_{2} s} \right]^{-1/s}
			\label{eqn:dpl}
		\end{equation}
		where $\alpha_{1} = 5/2$ and $\alpha_{2} = -(p-1)/2$ are the standard synchrotron 
		self-absorption and optically thin spectrum scenario, respectively, for the power law 
		distribution of electrons (Eqn. \ref{eqn:e_distribution}) with index $p$. 
		Further in this prescription $f_{a,0} \equiv f_{\nu}(\nu=\nu_{a}, t=t_0)$ i.e. spectral peak 
		is measured at the self-absorption frequency and at epoch $t=t_{0}$. 
		The parameter `$s$' in Eqn.\ref{eqn:dpl} is the smoothing parameter - higher value 
		of $s$ leads to sharper break compared to that with the lower $s$.
		
		The statistical uncertainties in fluxes measured and listed in Table \ref{tab:vla} are the image RMSs.
		Variations in the local weather at the telescope can introduce additional errors to 
		the measured image RMSs. Furthermore, calibration of the flux and phase calibrators
		can add systematic errors of a few percentage. Therefore, during the model fitting, 
		about $5\%$ of the total error was added in quadrature to account for these sources of errors. 
		
		\cite{Milisavljevic2013} has determined the epoch of explosion for SN\,2012au to be March 3.5 
		$\pm$ 2.0 based on the optical photometry.
		We used the same epoch of explosion for our fitting of the radio data. 
		The radio SED indicate that frequency $\nu_{m}< \nu_{a}$ during all the observations
		and in any case below the observing frequency $\nu_{obs}$.
		As a result, we can constrain the spectral break $\nu_{m} < 1$ GHz.
		
		We fit the multifrequency light curves of SN\, 2012au to the model described 
		in the appendix by using least square minimization 
		to estimate the self-absorption frequency ($\nu_{a}$) and spectral peak 
		($f_{\nu_{a}}$). 
		The best fit values turned out to be $f_{\nu_{a}} = 14.6 \pm 0.3$ mJy 
		at the self-absorption frequency $\nu_{a} = 6.0 \pm 0.1$ GHz
		at the epoch of $t_{0} = 25$ days after the burst and the resultant $\chi^{2}/{\rm dof} = 9$.
		We note that a significant contribution to the $\chi^{2}$ comes from 
		the observations at high-frequencies ($>$ 8 GHz).
		
		The shock wave expansion is best fit as $r=r_{0} (t/t_{0})^{0.93}$.
		For a compact, radiative envelope star, the outer density profile should tend to a power law, 
		$\rho_{sn} \propto r^{-k}$ with $k\sim 10.2$ \citep{Chevalier2006}.  For a wind medium, the shock
		front expands as $r\propto t^{m}$, with $m=(k-3)/(k-2)$, so $k=0.88$ is expected. Our estimate
		$m=0.93$ from the radio model, is therefore in a reasonable agreement.
		
		We left $p$ as well as `s' as free parameters to allow for the best match with 
		the radio spectra as well as light curves. 
		The fit converged to $p=2.7 \pm 0.1$ and $s=1.9\pm0.2$. The best fit value of $p$ 
		thus found is close to the typical values ($p\approx3$) found for SNe Ibc \citep{Chevalier2006}.
		
		At this stage we do not assume any particular value for $\epsilon_{e}$ or effectively for $\epsilon_{f}$. 
		Using our best fit spectral parameters for SN\,2012au and equations 
		\ref{eqn:beta0}-\ref{eqn:epsilonB0} we determined  
		$\beta_{0} = 0.20 ~\epsilon^{-1/17}_{f}$, $A_{\star} = 0.4 ~\epsilon^{5/17}_{f}$ 
		and $\epsilon_{B} = 0.3 ~\epsilon^{-13/17}_{f} $.
		
		The size of the supernova shock wave at epoch $t_{0} = 25$ days is thus $r_{0} = \beta_{0} c t_{0}/m$ 
		which from the derived quantities is estimated to be $r_{0} = 1.4 \times 10^{16} \epsilon^{-1/17}_{f}$ cm.
		As could be seen both the quantities $\beta$ and $r$ have a rather weak dependence 
		on the parameter $\epsilon_{f}$. 
		
		The value of $\epsilon_{B}$ could now be converted to the shock compressed magnetic 
		field $B = 0.54 \epsilon^{-4/17}_{f}$ G. Similarly, the minimum Lorentz factor of electron 
		energy distribution turns out to be $\gamma_{m} = 2.6 \epsilon^{2/17}_{f}$. This estimate
		is consistent with our assumption that the radiating electrons are relativistic, 
		with the spectrum extending down to $\gamma_{m} > 1$. Although $\gamma_{m}$ has 
		a weak dependence on $\epsilon_{f}$ a very low value of $\epsilon_{f}$ or 
		$\epsilon_{e}$ leading to $\gamma_{m} < 1$ would not be sensible.
		
		The mass loss rate of the progenitor star is defined as 
		$\dot{M} = 5 \times 10^{11} 4 \pi V_{w} A_{\star} ~\rm gm ~cm^{-1}$
		for the constant wind velocity $V_{w} = 1000$ km/s and normalization $A_{\star}$.
		For the derived value of $A_{\star}$ this gives, $\dot{M} = 
		3.6 \times 10^{-6} \epsilon^{5/17}_{f}~\rm M_{\odot} yr^{-1}$. 
		We note that this estimated mass loss rate is approximately a factor of 10 and 20 higher 
		than that of SN\,2002ap and SN\,1998bw, respectively. This value is still at the lower 
		end of the range of typical values found for the Wolf-Rayet stars 
		\citep{Cappa2004} which are considered to be the
		candidate progenitors of typical type Ib/c SNe as well as GRBs.
		
		The density profile of the circum-burst material is consistent 
		with a steady wind due to constant mass loss rate and constant wind velocity. 
		We can also estimate the number density 
		of the wind at a distance $r$ from the progenitor to be:
		\begin{equation}
			n_{e} = \frac{B^2_{0}}{8\pi} \frac{p-2}{p-1} \frac{\epsilon_f }{4  m_{e} c^{2} \gamma_{m,0}} \left( \frac{r}{r_{0}} \right)^{-2}
		\end{equation}
		For the measured physical parameters, we find $n_{e} \approx 9\times10^{2} ({r/r_{0}} )^{-2}\rm cm^{-3}$.
		
		Perhaps the most important physical parameter, the total internal energy for 
		the radio emitting material can now be estimated with 
		\begin{equation}
			E = \frac{4 \pi r^{3}_{0}}{\eta} \frac{1}{\epsilon_{B}} \frac{B^{2}_{0}}{8\pi} \left( \frac{r}{r_{0}} \right)^{-2},
		\end{equation}
		which yields a value of $E = 6 \times 10^{46} \epsilon^{2/17}_{f}$ erg at the epoch of $t_{0} = 25$ days.
		This estimated energy is intermediate between that of the two hypernovae - about 100 times lower 
		than SN\,1998bw and about 100 times higher than SN\,2002ap.
		
		Although the model in its current form does not take into account geometric effects, 
		such as collimation or asymmetry, it does provide reasonable fits to the observations 
		and large deviations from true physical parameters are unlikely. 
		
\section{Inverse Compton X-ray emission}
\label{sec:IC}

	The interaction of the SN shock with the circumstellar medium (CSM) is a well known source of X-ray radiation (\citealt{BjFr04}, \citealt{Chevalier2006}). Monitoring the high-energy emission from SNe therefore allows us to map the mass-loss history of the progenitor in the years before the explosion. In the case of hydrogen-stripped SNe exploding in low density environments, \cite{BjFr04} and \cite{Chevalier2006} demonstrated that the dominant X-ray emission mechanism during the first weeks to a month after the explosion is Inverse Compton (IC).  In this framework, X-ray photons originate from the up-scattering of optical photons from the SN photosphere by a population of electrons accelerated to relativistic speeds by the SN shock. 
	
	Following  \cite{Chevalier2006}, the IC X-ray luminosity depends on: (i) the density structure of the SN ejecta (ii) and of the CSM; (iii) the details of the electron distribution responsible for the up-scattering; (iv) the explosion parameters (ejecta mass $M_{\rm{ej}}$ and kinetic energy $E_{\rm{k}}$); and (v) the bolometric luminosity of the SN ($L_{\rm{IC}} \propto L_{\rm{bol}} $). 
	As inferred from the radio observations we assume wind medium 
	and an electron power-law index $p=2.7$. 
	From the modeling of the bolometric light-curve, \cite{Milisavljevic2013} constrained the ejecta mass and kinetic energy to be $M_{\rm{ej}}\sim 4\,M_{\sun}$ and $E_{\rm{k}}=10^{52}\,\rm{erg}$. Using these parameters and the formalism developed by \cite{Margutti12}, the \emph{Swit}-XRT observations contrain the mass-loss rate from the progenitor to $ \dot{M} \sim 5.6\times 10^{-6}\times \epsilon_{e,0.1}^{-2.39}\,\rm{M_{\sun}yr^{-1}}$, consistent with radio measurements, where $\epsilon_{e,0.1}$ is the fraction of energy into relativistic electrons in units of 0.1 and we assumed a wind velocity of $V_{w}=1000\,\rm{km\,s^{-1}}$. 
	
	One can further use independent estimates of mass loss rates from radio and X-ray emissions
	to constrain fractions of energy in accelerated electrons and magnetic field.
	As can be seen, $\dot{M}$ from X-rays depends strongly on $\epsilon_{e}$
	while that from radio, has a weak dependence on both $\epsilon_{e}$ 
	and $\epsilon_{B}$. Equating those mass-loss rates, we constrain 
	$\epsilon_{e} = 0.15 ~\epsilon^{0.11}_{B}$. 
	These quantities can be constrained rather strongly by alternative means 
	as shown in the next section.

\section{Constraints on $\epsilon_{e}$ and $\epsilon_{B}$}
\label{sec:constraints}
	In the previous section we used $\nu_{m} < 1$ GHz and, in the absence of forth 
	observable, expressed all the results 
	in terms of $\epsilon_{f}$. In this section we will relax the assumption about $\nu_{m}$
	and will try to constrain $\epsilon_{f}$.
	
\begin{figure}[h]
		\begin{center}
			\includegraphics[width=9cm]{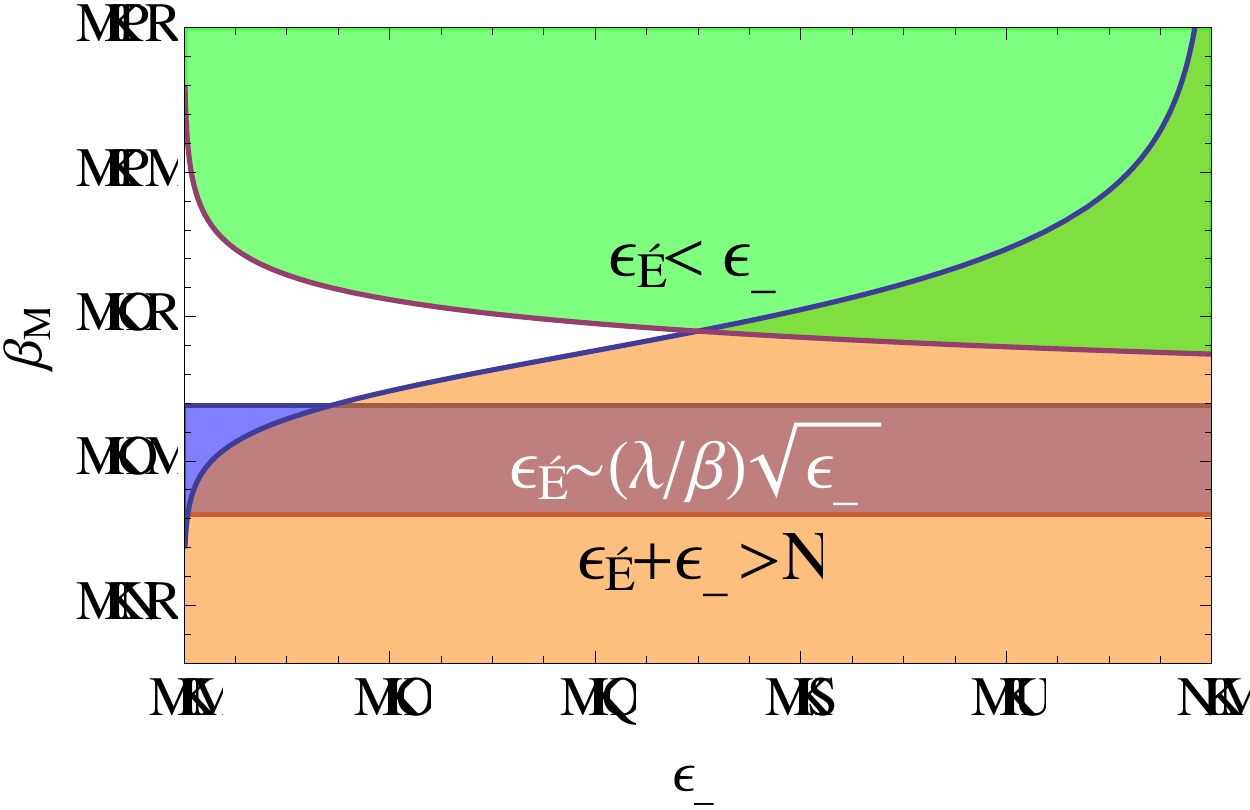}
			\caption{$\beta - \epsilon_{B}$ parameter space for SN\,2012au. 
			The green shaded region corresponds to $\epsilon_{e} < \epsilon_{B}$.
			The energy conservation requires that $\epsilon_{e} + \epsilon_{B} \leq 1$,
			ruling out the orange shaded region. The relation $\beta \propto \lambda^{-2/19}$
			enclosed between $0.5<\lambda<3.0$ is coded in blue.}
			\label{fig:constrain}
		\end{center}
	\end{figure}

	For simplicity, it is often assumed that $\epsilon_{f} = 1$ which translates to 
	$\epsilon_{e} = \epsilon_{B}$ or equipartition of energy between accelerated electrons 
	and shocked magnetic field. 
	In general, however, $\epsilon_{f}$ could have any non-zero positive value.
	By combining equations \ref{eqn:beta0} and \ref{eqn:epsilonB0}, we can eliminate $\nu_{m}$ between them
	and using the observed spectral parameter values for $\nu_{a}$ and $f_{\nu_{a}}$
	one obtains
	\begin{equation}
	\beta_{0} = \frac{0.24}{\epsilon^{1/21}_{B} \epsilon^{2/21}_{f}}
	\label{eqn:beta-eb}
	\end{equation}
	In Figure \ref{fig:constrain}, we show the $\beta-\epsilon_{B}$ phase-space
	with contours corresponding to two different values of $\epsilon_{f}$, viz.,
	$\epsilon_{f} = 1$ and $\epsilon_{e} + \epsilon_{B} = 1$. The restricted or 
	highly unlikely values are marked by shaded green and orange regions. 
	
	As discussed before $\epsilon_{f} = 1$ corresponds to the energy equipartition
	between shocked electrons and magnetic field. The green shaded region is 
	$\epsilon_{f} < 1$ or equivalently $\epsilon_{e} < \epsilon_{B}$. 
	Observations of SNe, supernova remnants and GRBs suggest that usually
	$\epsilon_{e} > \epsilon_{B}$. On the other hand, $\epsilon_{e} + \epsilon_{B} \leq 1$
	condition ought to be satisfied, leaving out the orange shaded region 
	as not-permissible.
	
	\citet{Medvedev2006} showed that $\epsilon_{e} = (\lambda/\beta) \sqrt{\epsilon_B}$
	for relativistic shocks such as in the case of GRBs. The dimensionless 
	parameter $\lambda$, which accounts for the geometry of the current filaments 
	and other plasma effects, has been found to range from about unity to a few based on
	the results of numerical simulations. 
	The results of \citet{Medvedev2006} should also hold for sub-relativistic shocks 
	with a moderate value of $\beta$, because the magnetic field in the circum-burst 
	medium ($\sim$ a few $\mu$G) is much smaller than the shock-generated fields 
	($\sim$ a few hundred mG) so that the shock structure is mediated by plasma 
	streaming instabilities. Using this relation
	along with Eqn.\ref{eqn:beta-eb} gives $\beta = 0.2 \lambda^{-2/19}$.
	The blue shaded region in Figure \ref{fig:constrain} bounds the higher and lower 
	limits obtained for $\lambda = 0.5$ and 3, respectively. 
	The lower value of $\lambda =0.5$
	limits the shock velocity to $\beta\leq0.22$, which along with the condition 
	$\epsilon_e + \epsilon_B \leq 1$ gives $\epsilon_{B} \leq 0.15$.
	
\section{Discussion}
\label{sec:discus}

	\subsection{Shock wave interaction with the steady wind}
	\label{sec:pisn}
		
		Towards the end of their lives very massive stars ($M\gtrsim70 M_{\odot}$) are likely 
		to undergo dynamical instability
		due to the $e^{-}-e^{+}$ pair formation leading to complete disruption of the star as 
		a pair-instability supernova (PISN) \citep{Barkat1967}. At a lower mass range of stars 
		favorable rotation periods and metallicities could result in episodic ejection of matter 
		shells from the outer stellar layers
		but not total disruption of the star. This is called the pulsational pair-instability 
		supernova (PPISN). Some of the most luminous supernovae are considered 
		to have originated as PPISN and PISN, for instance, SN\,2006gy 
		\citep{Smith2007,Ofek2007,Nomoto2007} and
		SN\,2007bi \citep{Gal-Yam2009,Young2010}.
		The stellar shell ejected as a result of the ultimate core collapse and SN explosion, 
		which would eventually take place in this modified environment, could subsequently 
		be expected to interact with the previously ejected shells, see e.g. \citet{Chatzopoulos2012}.
		Signatures of such interactions in the form of bumpy 
		light curves could serve as proof of a possible massive star progenitor as well as
		PPISN candidates.
		
		The case has been made for SN\,2012au as being a link between SLSN and normal 
		core collapse supernovae \citep{Milisavljevic2013}. The radio light curves of this event are smooth
		as shown in Figure \ref{fig:lc}. There is, therefore, no evidence of the shock 
		wave interacting with inhomogeneous circum-burst medium out 
		to $r \approx 10^{17}$ cm from the supernova site.
		The alternative interpretation that SN\,2012au was an asymmetric explosion 
		by \citet{Milisavljevic2013} based on spectral evolution and the two component velocity behavior 
		implied by several spectral features is, therefore, consistent with this finding.
		
		The lack of shock-shell interaction may also mean that there is a significant delay between
		the ejection of shells and the supernova thereby resulting in the shells being able to
		travel far away from the supernova site. Our observations suggest that the last few decades 
		of the progenitor of SN\,2012au were quiet during which no such shell ejections took place.
		
	\subsection{SN\,2012au in comparison with the hypernovae}
	\label{sec:hypern}
		
		The peak bolometric luminosity of SN\,2012au was about $L_{bol}\approx6 \times 10^{42}$ erg/s,
		modeling of which and the overall UV-optical-NIR band light curves of SN\,2012au suggest
		that the progenitor ejected $M_{ej} \approx 3-5 M_{\odot}$ with the kinetic energy of
		$E_{K}\approx 10^{52}$erg \citep{Milisavljevic2013, Takaki2013} as compared to the ordinary 
		type Ib/c SNe which have $E_{K}\approx 10^{51}$erg.
		This makes SN\,2012au a ``hypernova", a class which, as of now, constitutes only a few
		more peculiar SNe such as SN\, 1997ef, SN\,1998bw and SN\,2002ap, for instance. 
		While a few more candidates of this class have been identified and studied 
		in optical band, no detailed studies have been possible in the radio due to the paucity 
		of similar detected events in the local universe. The radio bright SN\,2012au
		therefore is a rare and important addition to this class.
		
		Tabel \ref{tab:hypernovae} compares important explosion and environmental parameters 
		of a few hypernovae estimated from their optical and radio evolution. SN\,1998bw was
		a supernova associated with GRB980425, and was extraordinary for several reasons:  
		broad-absorption features in the spectrum associated with velocities in excess of 30,000 km/s, 
		large kinetic energy ($> 10^{52}$ erg) in the ejecta
		radiating in optical (slower ejecta) and a relativistic shock wave in the front powered by 
		large kinetic energy ($> 10^{49}$ erg) leading to a bright radio emission. Similar broad
		absorption lines and large kinetic energy ($> 10^{52}$ erg) were inferred from 
		the optical emission of SN\,2002ap 
		evoking excitement in the community leading to intense observing campaigns.
		SN\,2002ap, however, turned out to be rather faint: about 1000 times fainter than its 
		predecessor SN\,1998bw, in radio emission. Detailed analysis of radio observations 
		suggested a low reservoir of energy ($E_{K} \sim 10^{45}$erg) 
		in the high velocity ejecta leading to the conclusion that broad-absorption lines
		do not necessarily serve as proxy for relativistic shock waves \citep{Berger2002}.
		
		Explosion parameters for SN\,2012au, inferred from optical observations are similar to those 
		of SN\,1998bw. The peak radio brightness was, however, intermediate
		- $\sim 10$ times fainter than SN\,1998bw but $>100$ times brighter than SN\,2002ap.
		While the shock wave is non-relativistic with $v\sim 0.2 c$, the kinetic energy is 
		significant, $E_{K} \sim 10^{47}$erg, which is about 100 times less than SN\,1998bw but 
		larger than that of SN\,2002ap by a similar factor (see Figure \ref{fig:EkGammaBeta}). 
		Thus, SN\,2012au occupies 
		the intermediate space between the extremes of SN\,1998bw and SN\,2002ap.
		
		The differences in the brightness of radio emission
		may be traced to the mass loss histories and velocities of the shock waves.
		The inferred mass-loss rate for SN\,2012au is at least an order of magnitude higher
		than both SN\,1998bw and SN\,2002ap. 
		We note that the metallicity of the explosion site of SN2012au is
		super solar, $Z~\sim 2 ~Z_{\odot}$. Given the mass loss and 
		metallicity relation $\dot{M} \sim Z^{0.8}$ \citep{Asplund2005}, we find that our measured 
		mass loss rate based on the detection of radio emission is in line with 
		the $\dot{M}$ values for Wolf-Rayet stars. The progenitors of SN\,1998bw and SN\,2002ap
		were massive stars with $M\sim 30-35$ and $20-25~M_{\odot}$, respectively, based on 
		the optical studies.  \cite{Milisavljevic2013} places loose limits on the progenitor of SN\,2012au 
		being $<80 ~M_{\odot}$.
		
		It is contrary, however, to our current understanding that such a massive progenitor 
		should be able to extract as much as $10^{52}$ erg or more 
		from the reservoir \citep{Burrows1998,Fryer1998}. More massive the core, 
		more effective is the neutrino cooling
		which drains the energy reservoir and effectively makes the explosion weaker.
		Under such a situation, as has been argued by several authors \citep{Bodenheimer1983,
		Hartmann1995,Woosley1999,Burrows2007}, we may have to abandon 
		the spherical explosion geometry resulting in the possibility of asymmetric explosions
		powered by the jets. The jet geometry and observer's orientation with respect 
		to the jet axis could be a plausible explanation for the diversity of SNe including 
		the large kinetic energies in optical but different radio emission properties.
		
	\subsection{Asymmetry in supernovae and SN\,2012au}
	\label{sec:asymme}
		
		Some of the key features in the optical spectra of SN\,2012au include the P-Cyg absorption profiles,
		asymmetries in the distribution of elements and their ions inferred from emission line profiles
		and the absence of Fe[III] in the iron plateau indicating high density of the Fe. \citet{Milisavljevic2013}
		attributes these features to the intrinsic explosion asymmetry based on the 
		comparisons with the model predictions by \citet{Chugai2000,Maeda2002,Maeda2003a,
		Maeda2003b,Maeda2006,Mazzali2004}. 
		These models have shown that aspherical explosions can significantly modify both, 
		the density profiles and distribution of elements within the ejecta.
		
		Among the global optical properties of SN\,2012au, the most distinct is 
		the slow spectroscopic and photometric evolution similar to that in SN\,1997dq 
		and SN\,1997ef \citep{Mazzali2004}, which were SN\,1998bw 
		like supernovae classified as ``hypernovae". 
		The spherical hydrodynamic models which describe the early light curves
		rather well of such energetic SNe produce fainter optical luminosities
		during the late linear decline phase. The two-component models
		which use a dense inner core in addition to the usual outer core reproduce 
		the late linear decline phase \citep{Chugai2000,Maeda2003a}.
		A similar explanation has been invoked by \cite{Milisavljevic2013} 
		in order to explain late-time emission properties of SN\,2012au.
		The dense core required to explain these properties
		is also a necessary feature of the jet driven asymmetric SNe models. 
		
		Although it would be difficult to find signatures of aspherical explosion in a SN, 
		the relative high velocity of the shock wave could be one of those as aspherical explosion
		is more likely to produce anisotropic velocity distribution of the shock waves than 
		a spherical explosion. 
		The velocity of the shock front that is powering the synchrotron radiation and thus 
		the radio brightness of SN\,2012au is estimated to be $v = 0.2\,c$. This places SN\,2012au
		among the faster shock fronts in supernovae. 
		
	\subsection{Energy as a function of velocity for GRBs and SNe Ibc}
	\label{sec:EkGammaBeta}
		
		Radio synchrotron emission in supernovae is generated primarily by the electrons heated 
		to relativistic temperatures behind the shock wave. Thus radio observations serve as 
		a probe of high velocity ejecta. Optical observations, on the other hand, probe
		the slowest ejecta in the explosion, which carry most of the mass and therefore 
		kinetic energy ($E_{K} = 0.3~M_{ej} v^{2}_{ej} \approx 10^{51}$ erg).
		
\begin{figure}
			\begin{center}
				\includegraphics[width=9cm,clip=False]{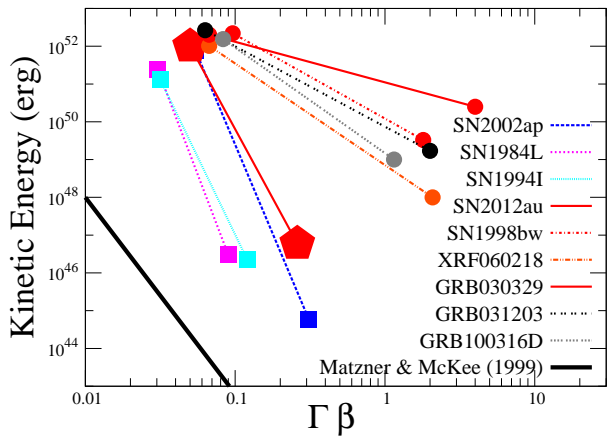}
				\caption{Energy as a function of velocity for GRBs and SNe Ib/c.
				The kinetic energy profile due to standard hydrodynamic collapse
				is shown as a solid black line. The high velocity ejecta 
				of SN\,2012au closely resembles the normal SNe, therefore marking it as 
				a standard event.
				\footnote{References: \citet{Schlegel1989,Swartz1991,Baron1993,Iwamoto1994,Young1995,
				Richmond1996,Iwamoto1998,Kulkarni1998,Hoflich1999,
				Matzner99,Millard1999,Woosley1999,Sollerman2000,Matheson2001,
				Nakamura2001,Berger2002,Mazzali2002,Berger2003a,Berger2003b,
				Mazzali2006,Soderberg2006,Milisavljevic2013,Margutti2013}}}
				\label{fig:EkGammaBeta}
			\end{center}
		\end{figure}

		In Figure \ref{fig:EkGammaBeta} we compare different types of stellar explosions:
		GRBs and supernovae Ib/c. Standard hydrodynamic collapse results in kinetic 
		energy profile $E_{K} \propto (\Gamma \beta)^{-5.2}$ shown by the solid black line
		\citep{Matzner99}. Therefore, much of the kinetic energy 
		is deposited in the slowly moving ejecta, leaving the high velocity or mildly relativistic ejecta
		with negligible amount of kinetic energy, in comparison. GRBs, however, have comparable 
		amounts of energy in slow as well as relativistic ejecta, which also points to the presence
		of a different source of energy - the central engine.
		
		The kinetic energy profile of SN\,2012au resembles closely to that expected from 
		the hydrodynamic collapse and to the absence of any central engine activity.
		Despite the faster shock wave, the radio emission from SN\,2012au conforms 
		to the simple picture of core collapse supernovae.
		
\section{Conclusions}
\label{sec:conclu}

	SN\,2012au displayed bright radio emission comparable to some of 
	the brightest core collapse SNe. It is also one of the most energetic
	supernova based on the energy estimates from optical luminosities
	and appears to form a link between SLSN and normal core-collapse
	supernovae. These properties make it an important SN to shed light 
	on the diversity among the core collapse SNe. 
	
	The smooth evolution of the radio brightness of SN\,2012au over a wide 
	range of frequency and time has been very well explained in terms
	of synchrotron emission from the shock that is interacting with
	the stellar wind of the progenitor. Based on the detailed analysis of 
	the radio emission we infer non-relativistic velocity for the shock
	wave of $v \approx 0.2\, c$ and a moderate mass loss rate
	of $\dot{M} = 3.6 \times 10^{-6}~\rm M_{\odot} ~yr^{-1}$
	for the constant wind velocity of 1000 km/s.
	We estimate the internal energy associated with the shock wave  
	to be $E\approx 10^{47}~\rm erg$.
	The light curves of SN\,2012au showed no sharp features or discontinuities, 
	indicating that the shock wave did not interact with any such shells 
	out to distance of $r \approx 10^{17}~\rm cm$ from the supernova site.
	Thus, the progenitor did not have such outbursts 
	during the last few decades before the supernova.
	
	The kinetic energy estimate from optical light curves 
	($E_{K} \approx 10^{52} ~\rm erg$) and expansion velocity makes SN\,2012au an 
	energetic event resembling SN\,1998bw and SN\,2002ap. 
	SN\,1998bw was associated with the GRB980425 and had the internal energy 
	budget of $10^{49}$ erg for the radio emitting relativistic ejecta. 
	SN\,2002ap on the other hand had a very low energy budget of only $10^{45}$ erg
	as inferred from the radio observations. SN\,2012au is intermediate - about 100 times 
	more energetic than SN\,2002ap and less energetic than SN\,1998bw 
	by the same factor. This difference in the energy budget may also be 
	the reason for the differences in luminosities.
	
	SN\, 2012au was an energetic supernova with a rare combination
	of properties shared by subsets of hydrogen poor hypernovae
	and super-luminous supernovae. These properties include
	high kinetic energies ($\sim 10^{52}$ erg), slow evolution,
	and late time optical spectral features potentially associated with
	explosion asymmetries. 
	
	We show that the radio emission from SN2012au is best explained by a 
	simple model of shock interaction with a steady wind.  We conclude that the
	unusual demise of the progenitor star was a result of the explosion properties.

\acknowledgments
	Support for this work was provided by the David and Lucile Packard Foundation
	Fellowship for Science and Engineering awarded to AMS.




\begin{deluxetable}{cccccccccccccccl}
\tablecaption{Model parameters of some hypernova candidates.\\
{\bf References :}
(1) \citet{Kulkarni1998}; (2) \citet{Li1999}; (3)\citet{Mazzali2001}; (4)\citet{Iwamoto1998};
(5) \citet{Berger2002}; (6) \citet{Mazzali2002};
(7) \citet{Iwamoto2000}; (8) \citet{Mazzali2000};
(9) \citet{Mazzali2004}; (10) \citet{Milisavljevic2013}; (11) this work}
\tablewidth{0pt}
\tablehead{ 
	SN		&$\nu_{p}$&$L_{\nu_{p}}$& $T_{p}$	& $\theta_{ep}$	&$r_{ep}$&$E_{ep}$&$B_{ep}$&$\gamma~\beta$&$\dot{M}$	&$T_{rise}$&$M_{Ni}$     &$M_{ej}$	  &$E_{k}$&Ref \\
	$\ldots$	& GHz	&erg/s-Hz		& days	& $\mu$~arc-sec&	cm	&erg		&	G    &$\ldots$		&$10^{-6}~M_{\odot}/yr$&days   	  &$M_{\odot}$&$M_{\odot}$&$10^{51}$erg	 &$\ldots$
}
\startdata
	1998bw	&6	& $7\times10^{28}$	&	16	&	100	&  $5.7\times10^{16}	$	& $10^{49}$ &0.4	&1.8		&0.25	&15.3	&  0.4	&	13	&	20	&  1,2,3,4\\
	2002ap  	&1.4	& $2\times10^{25}$  &	7	&	40	&  $5.4\times10^{15}$	& $10^{45}$ &0.2	& 0.31	&0.5	  	&\ldots	&  0.07	&	2.5-5	&	4-10	&  5,6\\
	1997ef  	&\ldots&\ldots 			&   \ldots 	&   \ldots 	&   \ldots	&	\ldots 			&	\ldots   &\ldots	&\ldots	&\ldots	&0.16	&	8	&	20	&	7,8\\
	1997dq  	&\ldots&\ldots 			&   \ldots 	&   \ldots 	&   \ldots	&   	\ldots   			&	\ldots   &\ldots	&\ldots	&\ldots	&0.16	&\ldots	&	20	&	9        \\
	2012au  	&8	& $7\times10^{27}$ 	&  20     	&	39	& $1.4\times10^{16}$	&  $10^{47}$&	0.4	&0.26	&3.6		&16.5	&0.3		&	3-5	&	10 	&	10,11
\enddata
\label{tab:hypernovae}
\end{deluxetable}


\appendix
\section{The Model}
	The supernova explosion drives a shock wave expanding into the surrounding medium 
	that has been modified by the wind driven by the exploding star over its lifetime. 
	The shock-wave sweeps the surrounding
	medium and heats it to relativistic temperatures and in the process converts the bulk
	kinetic energy of the incoming material into thermal energy of the shocked material.
	The electrons in the shock-heated plasma are accelerated in the post-shock 
	magnetic field to radiate synchrotron radiation.  Below we derive the
	expected flux density evolution. This model, which has been adopted 
	from the formalism of \citet{Chevalier1998,Soderberg2005,Chevalier2006},
	is presented here for the purpose of clarity and self-containedness.

	\subsection{The Electron energy distribution}
		We adopt a power law Lorentz factor distribution of shocked electrons with index $p$,
		as expected from the Fermi acceleration process in strong shocks:
		\begin{equation}
		n_{e}(\gamma_{e})~ d\gamma_{e} = K_{e}~ \gamma_{e}^{-p}~ d\gamma_{e}
		\label{eqn:e_distribution}
		\end{equation}
		for $\gamma_{m} \ll \gamma_{e} \ll \gamma_{u}$, where $\gamma_{m}$ and $\gamma_{u}$
		are the lower and upper energy cut-offs of the distribution and $\gamma_{e}$ the electron
		Lorentz factors. 
		The value of $\gamma_{m}$ and $K_{e}$ can then be found by assuming that the
		radiating electrons are concentrated in a thin static shell, have the non-thermal energy 
		distribution and that the amount of energy available for the electron acceleration is a constant
		fraction ($\epsilon_{e}$) of the total thermal energy density $U_{th}$.
		\begin{eqnarray}
			\int_{\gamma_{m}}^{\gamma_{u}} n_{e}(\gamma_{e})~d\gamma_{e} &=& 4 n \label{eqn:e_conserve} \label{eqn:e_density}\\
			\int_{\gamma_{m}}^{\gamma_{u}} \gamma_{e} ~m_{e} c^{2} ~n_{e}(\gamma_{e})~d\gamma_{e} &=& \epsilon_{e} U_{th}
			\label{eqn:e_energy}
		\end{eqnarray}
		Shock wave efficiently converts its kinetic energy 
		into internal energy of the shocked material. For a shock wave moving several times faster 
		than the local sound speed density compression of factor 4 is achieved and 
		the compressed medium trails the shock wave with speed lower by that factor.
		Further, it can be shown that the thermal energy density of the shocked medium 
		is given by $U_{th} = (9/8) n m_{p} \beta^{2} c^{2}$.  
		Using Eqn. \ref{eqn:e_distribution}, \ref{eqn:e_density},
		\ref{eqn:e_energy} and assuming $\gamma_{u} \gg \gamma_{m}$ one
		obtains
		\begin{eqnarray}
			\gamma_{m} & = & \epsilon_{e} \frac{9}{32}\frac{m_{p}}{m_{e}} ~\frac{p-2}{p-1}  \beta^{2} \label{eqn:gamma_m}\\
			K_{e} & = &4  n(p-1)\gamma_{m}^{p-1}    \label{eqn:Ke}
		\end{eqnarray}
		
	\subsection{Post-shock magnetic field}
		It is assumed that similar to $\epsilon_{e}$ a fraction $\epsilon_{B}$
		of the post-shock thermal energy goes into the magnetic field.
		\begin{eqnarray}
			\frac{B^{2}}{8\pi} &=& \epsilon_{B} U_{th}
			\label{eqn:B_at_rest}
		\end{eqnarray}
		Similar to $\epsilon_{e}$, in our entire discussion we will treat $\epsilon_{B}$
		as a constant in time.
		
	\subsection{Synchrotron Spectrum}
		It is assumed that the shocked electrons gyrate in the post-shock magnetic fields 
		and radiate synchrotron radiation. In order to characterize the spectrum
		we use the standard synchrotron formalism described in \citet{Rybicki1979}.
		The radio emission spectrum could be characterized by two break frequencies
		and the spectral peak. The two break frequencies, corresponding to the minimum 
		Lorentz factor of electrons ($\nu_{m}$) and synchrotron self-absorption ($\nu_{a}$),
		and the spectral peak ($F_{\nu_{m}}$) are derived below.
		
	\subsubsection{Minimum Lorentz factor $\gamma_{m}$ and the corresponding spectral break}
		The power-law distribution of the electrons has a lower Lorentz factor cut-off
		which we identify as $\gamma_{m}$ (Equation~\ref{eqn:gamma_m}).
		We adopt expression by ~\citet{Rybicki1979} for the characteristic synchrotron 
		frequency corresponding to $\gamma_{m}$,
		
		\begin{equation}
			\nu_{m} = \frac{3}{4\pi}\,\frac{\gamma_{m}^{2} q B}{m_{e} c}
			\label{eqn:nu_m_at_rest}
		\end{equation}
		where $q$ is the electric charge, $m_{e}$ is the mass of electron. 
		
	\subsubsection{Spectral break due to synchrotron self absorption}
		By approximating the thickness of the shocked radiating plasma to be $dr = r/\eta$ 
		the optical depth can be approximated as $\tau_{\nu} \sim 4 \alpha_{\nu} ~dr$. 
		For the synchrotron self-absorption coefficient $\alpha_{\nu}$, we used equation 6.53 of 
		\citet{Rybicki1979} and inverted the relation $\tau_{\nu} = 1.0$ to obtain the self-absorption frequency 
		\begin{equation}
			\nu_{a} = \frac{1}{\text{$m_{e}c$}} 
			   2^{\frac{2-p}{p+4}} 3^{\frac{p+1}{p+4}} \pi ^{-\frac{p+2}{p+4}} q^{\frac{p+6}{p+4}} 
			   (p-1)^{\frac{2}{p+4}}
			   B^{\frac{p+2}{p+4}} n^{\frac{2}{p+4}}
			   \text{$\gamma_{m}$}^{\frac{2 (p-1)}{p+4}} 
			    (r/\eta)^{\frac{2}{p+4}} 
			   \Gamma\left(x_{1}\right)^{\frac{2}{p+4}} 
			   \Gamma \left(x_{2}\right)^{\frac{2}{p+4}}
			\label{eqn:nu_a_at_rest}
		\end{equation}
		where $\Gamma(x_{1})$ and $\Gamma(x_{2})$ are the Gamma functions with 
		$x_{1} = (p/4+1/6)$ and $x_{2} = (p/4+11/6)$.
		
	\subsubsection{Spectral Brightness}
		The observed flux at a luminosity distance $d_{L}$ from a source of specific 
		intensity $I_{\nu}$ is obtained by integrating over all solid angles: 
		$f_{\nu} = \int I_{\nu}(\theta) cos \theta ~d\Omega$ and could be well approximated 
		as $f_{\nu} \simeq (j_{\nu}/\alpha_{\nu}) \Omega$ where $j_{\nu} = P_{\nu}/4 \pi$
		 is the emissivity of an isotropic emitter and $\alpha_{\nu}$ is the synchrotron self-absorption 
		 coefficient. This gives
		\begin{equation} 
			f_{\nu} \simeq \frac{1}{4 \pi} \left( \frac{P_{\nu}}{\alpha_{\nu}} \right) \Omega
		\end{equation}
		
		Following \citet{Rybicki1979} the synchrotron power per unit frequency emitted by a relativistic 
		electron is given by
		\begin{equation} 
			P_{\nu}(\nu,\gamma) = \frac{ \sqrt{3}}{2 \pi} \frac{q^{3} B}{m_{e} c^{2}} 
			F \left[ \frac{\nu}{\nu_{crit}(B,\gamma)}  \right]
		\end{equation}
		where critical frequency is defined as 
		\begin{equation}
			\nu_{crit} = \frac{3x_{p}}{4\pi}\,\frac{\gamma_{m}^{2} q B}{m_{e} c}
		\end{equation}
		and $F(x)$ is the synchrotron function discussed in \citet{Rybicki1979} [see Figure 6 
		and Equations 6.34 of \citealt{Rybicki1979}].
		For a power-law Lorentz factor distribution of electrons, Eqn. \ref{eqn:e_distribution},
		power per unit volume per unit frequency $P_{\nu}(\nu,\gamma)$ is given by
		\begin{equation}
			P_{\nu}(\nu,\gamma) = \frac{ \sqrt{3}}{2 \pi} \frac{q^{3} K_{e} B}{m_{e} c^{2} (p+1)} 
			\Gamma(x_{3}) \Gamma(x_{4}) \left( \frac {2 \pi ~m_e c ~\nu}{3 q B} \right)^{-(p-1)/2}
			\label{eqn:synch-power}
		\end{equation}
		and $x_{3} = (p/4 + 19/12), x_{4} = (p/4-1/12)$ [Equation 6.36 of \citealt{Rybicki1979}].
		
		\subsection{Spectral Evolution}
		The density profile of the surrounding medium which the shock-wave is ploughing
		through plays an important role in dictating the dynamics of its evolution and
		subsequently that of the radiation spectrum. There is a strong evidence that
		massive stars are the progenitors of radio SNe of Type Ib/c \citet{}. 
		Therefore the SN generated shock-wave should be expanding into the stellar wind 
		of the progenitor star ($\rho=A r^{-2}$). Considering this we calculate the spectral and temporal evolution
		in wind circumburst environment below which can be easily 
		generalized to a different density profile. In what follows, we parameterize the wind
		density profile by $A=\dot{M}/4 \pi V_{w} = 5\times 10^{11} A_{\star} ~\rm g~cm^{-1}$
		using the mass loss rate $\dot{M}=10^{-5} ~\rm M_{\odot}/yr$ and wind velocity 
		$V_{w}=1000 ~\rm km/s$ \citep{Chevalier1999}.
		 
		The instantaneous synchrotron spectrum can then be characterized 
		by two break frequencies ($\nu_{a}$ and $\nu_{m}$) and the spectral peak $F_{m}$.
		Usually, however, the peak frequency $\nu_{m}$ is below observing band and therefore 
		the spectral peak is rarely observed in SNe. We therefore express it in terms of
		an observable $f_{\nu_{a}}$ or the flux at the self-absorption frequency which is 
		related to the spectral peak $f_{\nu_{a}} = F_{m} (\nu_{a}/\nu_{m})^{-(p-1)/2}$.
		Because the shock wave, and therefore the radiating plasma behind it, is expected
		to be moving at non-relativistic or mildly relativistic velocities we have
		neglected relativistic effects including doppler boosting, for simplicity.
		Equations \ref{eqn:gamma_m}-\ref{eqn:synch-power} then give
		\begin{eqnarray}
			\nu_{m}	&	\approx	&
				7\times10^{4} q \frac{m^{2}_{p}}{m^{3}_{e}}
				\left(\frac{p-2}{p-1}\right)^{2} \sqrt{A_{\star}} \sqrt{\epsilon_{B}} \epsilon^{2}_{e}
				\frac{\beta^{5}}{r} {\rm Hz} \label{eqn:SpecPara1}\\
			f_{{\nu_{a}}}	&	\approx	&8 \times 10^{11} 
				q^{\frac{p (1.1 p+62.6)+89.8}{p+4}-\frac{1}{2}} c^{\frac{5(p/2+1)}{p+4}}
			   	m_{p}^{\frac{5 p-5}{p+4}} m_{e}^{\frac{11\, -3.5 p}{p+4}}
				~f_{p} f_{\Gamma}A_{\star}^{\frac{p+6.5}{p+4}} 
				\beta^{\frac{12 p-7}{p+4}} \eta^{-\frac{5}{p+4}} 
				\epsilon^{\frac{p+1.5}{p+4}}_{B} \epsilon^{\frac{5 p-5}{p+4}}_{e}
			      d_{L,100}^{-2} \,{\rm mJy} \label{eqn:SpecPara2}
		\end{eqnarray}
		along with the synchrotron self absorption frequency
		\begin{equation}
			\nu_{a} =  5 \times 10^{40} ~q^{\frac{38.8+28.5p}{p+4}} c^{\frac{p+2}{p+4}} \left( \frac{m_{e}}{m_{p}} \right)^{\frac{-2(p-1)}{p+4}}
				\left[ \frac{p-2}{p-1} \right]^{\frac{2(p-1)}{p+4}} (p-1)^{\frac{2}{p+4}} 
				A^{\frac{3+p/2}{p+4}}_{\star} 
				\frac{\beta^{\frac{5p-2}{p+4}}}{r} 
				\epsilon^{\frac{p/2+1}{p+4}}_{B} \epsilon^{\frac{2(p-1)}{p+4}}_{e}
				\eta^{-\frac{2}{p+4}}_{12}  ~\Gamma^{\frac{2}{p+4}}(x_{3}) ~~\Gamma^{\frac{2}{p+4}}(x_{4})
				~{\rm Hz}
			\label{eqn:SpecPara3}
		\end{equation}
		
		where $f_{p}=\left(\frac{p-2}{p-1}\right)^{\frac{5. p-5.}{p+4.}} \frac{(p-1)^{\frac{5.}{p+4.}}}{p+1}$ and 
		$f_{\Gamma} = \Gamma(x_{1}) \Gamma(x_{2}) \Gamma^{-\frac{p-1}{p+4}}(x_{3}) \Gamma^{-\frac{p-1}{p+4}}(x_{4})$,
		and $d_{L,100}$ to be the luminosity distance in units of 100 Mpc.

\end{document}